\documentclass[aps,prd,groupedaddress,showpacs,showkeys,amsmath,amssymb]{revtex4}
\usepackage{verbatim,graphicx,amssymb,amsbsy,bm,amsmath,rotating,epsfig}
\usepackage{psfrag}
\usepackage{hyperref}
\newcommand{\be}{\begin{equation}}
\newcommand{\ee}{\end{equation}}
\newcommand{\bea}{\begin{eqnarray}}
\newcommand{\eea}{\end{eqnarray}}
\begin{document}
\title{Role of Sterile Neutrino Warm Dark Matter in Rhenium and Tritium Beta Decays}
\author{H.J. de Vega}
\affiliation{LPTHE Universit\'e Pierre et Marie Curie (Paris VI),
Laboratoire Associ\'e au CNRS UMR 7589, Tour 24, 5eme. \'etage, Boite 126, Place Jussieu, 75252 Paris, cedex 05, France}
\affiliation{Observatoire de Paris, LERMA. Laboratoire 
Associ\'e au CNRS UMR 8112. \\ 61, Avenue de l'Observatoire, 75014 Paris, France.}
\author{O. Moreno}
\affiliation{Departamento de F\'{\i}sica At\'omica, Molecular y Nuclear,   
Facultad de Ciencias F\'{\i}sicas, Universidad Complutense, 28040 Madrid, Spain}
\author{E. Moya de Guerra}
\affiliation{Departamento de F\'{\i}sica At\'omica, Molecular y Nuclear,   
Facultad de Ciencias F\'{\i}sicas, Universidad Complutense, 28040 Madrid, Spain}
\author{M. Ram\'on Medrano}
\affiliation{Departamento de F\'{\i}sica Te\'orica I,
	        Facultad de Ciencias F\'{\i}sicas,
                Universidad Complutense,
                28040 Madrid, Spain}
\author{N. G. S\'anchez}
\affiliation{Observatoire de Paris, LERMA. Laboratoire 
Associ\'e au CNRS UMR 8112. \\ 61, Avenue de l'Observatoire, 75014 Paris, France.}
\date{\today}

\begin{abstract}
  Sterile neutrinos with mass in the range of one to a few keV are
  important as extensions of the Standard Model of particle physics
  and are serious dark matter (DM) candidates. This DM mass scale
  (warm DM) is in agreement with both cosmological and galactic
  observations. We study the role of a keV sterile
  neutrino through its mixing with a light active neutrino in Rhenium
  187 and Tritium beta decays. We pinpoint the energy spectrum of the
  beta particle, $ 0 \lesssim T_e \lesssim (Q_{\beta} - m_s) $, as the
  region where a sterile neutrino could be detected and where its mass
  $m_s$ could be measured. This energy region is at least 1 keV away
  from the region suitable to measure the mass of the light active
  neutrino, located near the endpoint $ Q_{\beta} $. The emission of a
  keV sterile neutrino in a beta decay could show up as a small kink
  in the spectrum of the emitted beta particle. With this in view, we
  perform a careful calculation of the Rhenium and Tritium beta
  spectra and estimate the size of this perturbation by means of the
  dimensionless ratio ${\mathcal R}$ of the sterile neutrino to the
  active neutrino contributions. We comment on the possibility of
  searching for sterile neutrino signatures in two experiments which
  are currently running at present, MARE and KATRIN, focused on the
  Rhenium 187 and Tritium beta decays respectively.

\end{abstract}
\maketitle

\section{Introduction \label{sec:intro}}

It is well known that dark matter (DM) is not described by the
Standard Model (SM) of particle physics. Many extensions can be
envisaged to include DM particles, coupled weakly enough to the SM
particles to fulfill all particle experimental constraints, namely the
fact that DM has not been detected so far in any particle physics
experiment. On the other hand, cosmological and astrophysical
constraints such as the ones coming from the dark matter density and
the galaxy phase space density, or alternatively, the universal galaxy
surface density, lead to DM candidates in the keV mass scale, namely
warm DM (WDM), refs. \cite{pri,wdm,wdm2,psd,sup,dk,bb}. A keV mass scale
sterile neutrino is the front running candidate for WDM. Other
possible WDM candidates in the keV mass scale are gravitinos, light
neutralinos and majorons \cite{pri,ste}.

Considering the first WDM candidate, sterile neutrinos can be
naturally embedded in the SM of particle physics. They do not
participate in weak interactions, and hence they are singlets of
color, weak SU(2) and weak hypercharge. One sterile neutrino per
lepton family could be expected, of which the lightest one
(i.e. electron family) would have a lifetime of the order of the
Hubble time and could be considered a DM candidate.

In this work, we consider the role played by a 1-2 keV sterile
neutrino in Rhenium 187 and Tritium beta decay experiments. The
left-handed neutrino flavor state $ \nu_e $ (and equivalently for $
\bar{\nu}_e $) will be a mixing of two mass eigenstates: one light
active neutrino mass state ($ \nu_l $) and one keV scale sterile
neutrino mass state ($ \nu_s $). Other neutrino mass states will not
be taken into account for the time being. The mass $ m_l $ of the
lightest active neutrino state is negligible ($ m_l \ll $ eV ) in
comparison with the mass $ m_s $ of the keV sterile mass state. The
smallness of the mixing angle $ \zeta $ makes sterile neutrinos
difficult to detect.

Sterile neutrinos in the beta decay of Rhenium 187 are currently
searched for by the Microcalorimeter Arrays for a Rhenium Experiment
(MARE) \cite{mar}. In this decay the available energy is $
Q_{\beta}(^{187}\text{Re}) \simeq 2.469 $ keV. The beta decay of
Rhenium 187 into Osmium 187 is a first forbidden unique Gamow-Teller
process ($5/2^{+} \rightarrow 1/2^{-}$).

Up to now, the non observation of keV scale sterile neutrinos in the
beta decay of Rhenium 187 gave an upper bound on the mixing angle $
\zeta < 0.095 $ for 1 keV steriles \cite{gal}, which is compatible
with the cosmological constraints on the mixing angle, $ \zeta <
10^{-3} $, appropriate to produce enough sterile neutrinos to account
for the observed DM. However, the amount of the sterile neutrinos that
could be produced in the early universe also depends on the production
mechanism, which is model dependent. We refer for that to the original
references \cite{wdm2}.

The Karlsruhe Tritium Neutrino Experiment (KATRIN) is currently
studying the Tritium beta decay \cite{kat} and, if suitably adapted,
it could study the presence of a sterile neutrino as well. In this
decay the available energy is $Q_{\beta}(^{3}\text{H}_1) \simeq 18.6 $
keV. The beta decay of Tritium into Helium 3 is an allowed transition
($1/2^{+} \rightarrow 1/2^{+}$) with Fermi and Gamow-Teller
contributions.  Clearly, KATRIN has in principle the potential to
detect sterile neutrinos with mass up to 18 keV.  However, the main
difficulty in detecting WDM sterile neutrinos comes from the smallness of
the mixing angle between the active and sterile neutrino $ \zeta <
10^{-3} $. Such range of values for $ \zeta $ are too small for the
present experimental sensitivities \cite{kat,cw} and would require a
source with a large stability to reduce the systematic errors.

Detection of massive neutrinos by $\beta$-decay has been proposed in
Ref. \cite{shr}. Other methods proposed to detect sterile neutrinos
include measurements of the nuclear recoil \cite{ponte,recule} and sterile
neutrino capture on $\beta$-decaying nuclei \cite{chi}.

In 1985 evidence for the emission of a 17 keV mass neutrino in Tritium
beta decay was reported by J. J. Simpson \cite{sim}. The evidence was
hotly debated, new experiments gave clear negative results and by 1993
the general conclusion was reached that there are no 17 keV neutrinos
\cite{wn,af}. The experiments in the nineties using $^{63}$Ni,
$^{35}$S and other nuclei yielded an upper bound $ \zeta < 0.03 $
\cite{wn}. This bound is not restrictive for DM because the
cosmological constraints based on the observed average DM density
indicate for the currently popular models of DM sterile neutrinos a
much lower bound, $\zeta < 10^{-3}$ \cite{dk,bb}.

Sections II and III deal with keV dark matter from the cosmological
and galactic point of view. WDM (DM particle mass between 1 keV and 10
keV, and decoupling temperature $ T_d \sim 100 $ GeV) produces the
observed small (galactic) structures, as well as the large scale and
cosmological structures, the observed cored density profiles and the
right surface density value, while GeV WIMPS ($ m \sim 100 $ GeV, and
$ T_d \sim 5 $ GeV, cold DM) inevitably produce a host of small-scale
structures and cusped profiles which have not been observed, as well
as a galaxy surface density much higher than observed. This summarizes
our motivation for proposing a laboratory search for sterile neutrinos
as DM candidates.

In Section IV we analyze the role that sterile neutrinos would play in
the electron spectrum of Rhenium beta decay taking into account
contributions from the electron $s$- and $p$-waves \cite{dvor,AR}.
The electron kinetic energy range $T_e$ suitable for the detection of
sterile neutrinos lies between 0 and $ (Q_{\beta}- m_s)$, where $ m_s
$ is the mass of the keV sterile neutrino. On the contrary, the
electron kinetic energy region close to the endpoint energy $
Q_{\beta} $ is the one suitable for the detection of light active
neutrinos. Systematic uncertainties such as Beta Environmental Fine
Structure (BEFS) are not considered here \cite{mar}. In order to
analyze the sterile neutrino effect, we introduce the dimensionless
ratio ${\mathcal R}$ of the sterile neutrino contribution to the
active neutrino contribution. It allows us to compare two regions of
the same spectrum: the one where the keV neutrino imprints a kink on
the spectrum, and the one near the endpoint where the active light
neutrino effect shows up. The Kurie function is also analyzed and
expressed in terms of the ratio ${\mathcal R}$.

In Section V we study the role of sterile neutrinos in Tritium decay
\cite{kat}, where the emitted electrons are purely
$s$-wave. Analogously to the Rhenium beta decay, the kinetic energy
region relevant for the sterile neutrino detection is the low energy
range $ 0 \leq T_e \leq (Q_{\beta}- m_s) $, while for the active
neutrino it is the one close to the endpoint energy $ Q_{\beta} $.

Finally, in Section VI we present our conclusions. Natural units $
\hbar = c = 1 $ are used all over this paper.

\section{Dark matter \label{sec:DM}}

Although dark matter was noticed seventy-five years ago \cite{zw,oo},
its nature is not yet known.  Dark matter (DM) is needed to explain
the observed structures in the Universe, in particular galaxies. DM
particles must have been non-relativistic by the time of structure formation
in order to reproduce the observed small structure at $ \sim 2-3 $
kpc.

The connection between the scale of the formed structure and the mass
of the DM particle follows from the value of the free-streaming length
$ l_{fs} $ \cite{kt}. This is the distance that the DM particles can
freely travel. Structures at scales smaller than $ l_{fs} $ are erased
by free-streaming and hence $ l_{fs} $ provides a lower bound on the
size of DM dominated structures.  WDM particles with mass in the keV
scale give $ l_{fs} \sim 100 $ kpc while 100 GeV cold dark matter
(CDM) particles produce an extremely small $ l_{fs} \sim 0.1 $ pc. A $
l_{fs} \sim 100 $ kpc is in nice agreement with the astronomical
observations of galaxies \cite{gil} (smaller objects like stars are
made up of baryons, not of DM), as well as at cosmological scales.

The GeV CDM free-streaming length $ l_{fs} $ is a million times
smaller and would lead to the existence of a host of CDM smaller scale
structures till the size of the solar system.  No structure of such
type has ever been observed.  Lighter DM particles in the eV scale
(hot dark matter, HDM) have a free-streaming length $ l_{fs} \sim $
Mpc and hence would erase all existing structures below the Mpc scale
in contradiction with all observations.  This is why HDM has been
ruled out \cite{dod}.

The reason why CDM does not work is simple: CDM particles in the GeV
scale are too slow (too cold), which prevents them to erase the small
scale structure, while the eV particles (HDM) are excessively fast,
which erases all structures. In between, WDM keV particles are able to
produce the observed structures.

Astronomical observations strongly indicate that dark matter halos
have cored profiles till scales below 1 kpc.  On the contrary, CDM
simulations (particles heavier than 1 GeV) always give cusped
profiles. No cusped profiles have been ever observed. Linear profiles
computed from the Boltzmann-Vlasov equation turn out to be cored for
WDM and cusped for CDM indicating that WDM does reproduce the
astronomical observations \cite{sup}.

The surface density in DM-dominated galaxies is defined by $ \mu_0
\equiv \rho_0 \; r_0 $ where $ \rho_0 $ is the central core density
and $ r_0 $ is the core radius. $ \mu_0 $ turns out to be universal,
taking the same value up to $ \pm 10\% $ for galaxies of different
sizes, morphologies, Hubble types and luminosities \cite{gd}.  The
surface density value predicted by CDM simulations is 1000 times
larger than the observed value \cite{hof}, while the surface density
for keV WDM computed from the Boltzmann-Vlasov equation is in full
agreement with the observed value of $ 120 \; ({\rm MeV})^3 $,
indicating again that WDM does reproduce the astronomical observations
\cite{sup}.

Constraints of the DM particle mass to the keV range are obtained from
combining theoretical analyses with the observed values of dark matter
densities and phase space densities today (density over the cube of
the velocity dispersion) of dwarf spheroidal galaxies.

Recent radioastronomy observations of velocity widths in galaxies from
21cm HI surveys clearly favours WDM over CDM \cite{pz}. WDM
simulations contrasted to astronomical observations suggest a WDM
particle mass slightly above 1 keV.  Constraints from large scale
structure give this value too \cite{cata}.  Recent cosmological WDM
N-body simulations with keV sterile neutrino WDM clearly show the
agreement of the predicted small scale structures with the
observations, while CDM simulations do not agree with observations at
such scales \cite{kama}.

None of the predictions of CDM simulations at small scales (cusps,
substructures, dark disks, ...) have been observed. Here are some
examples. The CDM satellite problem, namely that CDM simulations
predict too many satellites in the Milky Way and only 1/3 of
satellites predicted by CDM simulations around our galaxy are
observed. The surface density problem, which consists of the galaxy
surface density for CDM simulations being 1000 larger than observed
\cite{sup,hof}. And the voids problem and the size problem, that have
to do with the fact that CDM simulations do not produce big enough
galaxies \cite{tikho,meu,cha}. Further WDM properties are discussed
in \cite{mas}.

Notice that all DM observable effects discussed above only arise from
the gravitational behaviour of the DM. Galaxy properties are
independent of the non-gravitational couplings of
the DM particles, provided that their couplings are small enough.

DM may decouple at or out of thermal equilibrium. The distribution
function freezes out at decoupling.  Whether they decouple at or out
of equilibrium depends on the non-gravitational couplings of the DM
particle.  Normally, sterile neutrinos are so weakly coupled that they
decouple out of thermal equilibrium.  The functional form of the DM
distribution function depends on the DM particle couplings and is
therefore model dependent.

Sterile neutrinos can decay into an active-like neutrino and a
monochromatic X-ray photon with an energy half the mass of the sterile
neutrino. Observing the X-ray photon provides a way to observe sterile
neutrinos in DM halos \cite{raX,raX2}.

WDM keV sterile neutrinos can be copiously produced in the supernovae
cores.  Supernovae (SN) stringently constrain the neutrino mixing
angle squared to be $ \lesssim 10^{-9} $ for $ m > 100$ keV, in order
to avoid excessive energy lost. However, for smaller masses the SN bound is
not so direct. Within the models worked out till now, mixing angles
are essentially unconstrained by SN in the keV mass range \cite{raf}.

Sterile neutrinos are produced out of thermal equilibrium and their
production can be non-resonant, in the absence of lepton asymmetries,
or resonantly enhanced, if lepton asymmetries are present.  keV
sterile neutrino WDM in minimal extensions of the Standard Model is
consistent with Lyman-alpha constraints within a wide range of the
model parameters.  Lyman-alpha observations give a lower bound for the
sterile neutrino mass of 4 keV only for sterile neutrinos produced in
the case of a non-resonant (Dodelson-Widrow) mechanism
\cite{viel,rusos}. The Lyman-alpha lower bounds for the WDM particle
mass are smaller in the Neutrino Minimal Standard Model, where sterile
neutrinos are produced by the decay of a heavy neutral scalar, and for
fermions in thermal equilibrium. Moreover, the number of observed
Milky-Way satellites indicates lower bounds between 2 and 13 keV for
different models of sterile neutrinos.

In summary, contrary to CDM, WDM essentially  {\it works}, reproducing in a
natural way the astronomical observations of structures over all
scales, small as well as large and cosmological scales.  The sterile
neutrino with mass in the keV scale appears as a serious candidate for
WDM.  Galaxy observations alone cannot determine the DM particle
properties other than the mass and the decoupling temperature. A
direct particle detection is necessary to pinpoint and determine which
particle candidate describes DM. Beta decay is a promising way to
detect DM sterile neutrinos.

\section{Dark matter and keV sterile neutrinos}

As it is known, DM is not described by the Standard Model (SM) of
particle physics. However, many extensions of the SM can be envisaged
to include a DM particle with mass in the keV scale and coupled weakly
enough to the Standard Model particles so as to fulfill all particle
physics experimental constraints, coming mainly from the fact that DM
has not been detected so far in any particle physics
experiment. Besides sterile neutrinos, possible DM candidates in the
keV mass scale are gravitinos, light neutralinos, majorons,
etc. \cite{ste}.

As particle physics motivations for sterile neutrinos one can advance
that there are both left- and right- handed quarks (with respect to
chirality) while active neutrinos are only left-handed.  It is thus
natural to have right-handed neutrinos $ \nu_R $ besides the known
left-handed active neutrinos. This argument is called `quark-lepton
similarity'.

Sterile neutrinos can be naturally embedded in the SM of particle
physics with the symmetry group $ SU(3)_{color} \otimes SU(2)_{weak}
\otimes U(1)_{weak ~ hypercharge} $. Leptons are singlets under color
SU(3) and doublets under weak SU(2) in the SM. Sterile neutrinos $
\nu_R $ do not participate in weak interactions. Hence, they must be
singlets of color SU(3), weak SU(2) and weak hypercharge U(1).

Let us consider a simple embedding of the sterile neutrino in the
Standard Model.  More elaborated sterile neutrino models have been put
forward \cite{modest}. The SM Higgs $ \Phi $ is a SU(2) doublet with a
nonzero vacuum expectation value $ \Phi_0 $. This allows a Yukawa-type
coupling with the left- and right-handed leptons:
\begin{equation}\label{lagneu}
L_{Yuk} = y \; {\bar \nu_L} \; \nu_R \; \Phi_0 + h. c. \: ,
\end{equation}
where $y$ is the Yukawa coupling, and
\begin{equation}\label{phi_0}
\Phi_0 = \left( \begin{array}{c} 0 \\  v \\  \end{array}  \right)
\quad , \quad v = 174 \; {\rm GeV}.
\end{equation}
These terms in the Lagrangian induce a mixing (bilinear) term between
$ \nu_L $ and $ \nu_R $ allowing for transmutations $ \nu_L
\Leftrightarrow \nu_R $.  Mixing and oscillations of particle states
are typical of low energy particle physics. Further well known
examples are: (i) flavor mixing: e-$\mu$ neutrino oscillations, which
explain solar neutrinos, (ii) $ K^0-{\overline K}^0, B^0-{\overline
  B}^0 $ and $ D^0-{\overline D}^0 $ meson oscillations in connection
with CP-violation.

As a consequence of the Lagrangian in Eq. (\ref{lagneu}), the neutrino
mass matrix takes the form
\begin{equation}\label{bilinear}
\left( {\bar \nu_L} \; {\bar \nu_R}\right) \; 
\left( \begin{array}{cc}  0  &  m_D \\   m_D  & M  \\   \end{array}  \right)  
\; \left( \begin{array}{c} \nu_L \\  \nu_R \\  \end{array}  \right)
\end{equation}
where $ M $ is the mass term of the right-handed neutrino $ \nu_R $,
and $m_D = y \; v $ with $ M \gg m_D $.

The masses of the active and sterile neutrinos are given by the seesaw
mechanism. The mass eigenvalues in this simple model take the form: $
m_D^2/M $ (active neutrino) and $ M $ (sterile neutrino), with
eigenvectors $\nu_{active} \simeq \nu_L - \frac{m_D}{M} \; \nu_R$
(active neutrino) and $\nu_{sterile} \simeq \nu_R + \frac{m_D}{M} \;
\nu_L , \quad M \gg m_D^2/M$ (sterile neutrino). Choosing $M \sim 1$
keV and $ m_D \sim 0.1 $ eV yields $ m_D^2/M $ about $ 10^{-5} $ eV,
consistent with observations. This corresponds to a mixing angle $
\zeta \sim m_D/M $ about $ 10^{-4} $ and would be appropriate to
produce enough sterile neutrinos to account for the observed
DM. However, notice that the amount of the sterile neutrinos produced
in the early universe also depends on the production mechanism, which
is model dependent.  The smallness of the mixing angle $ \zeta $ makes
sterile neutrinos difficult to detect.

One sterile neutrino per lepton family could be expected, of which the
lightest one (i.e. electron family) would have a lifetime of the order
of the Hubble time and could be considered a DM candidate. In summary,
the empty slot of right-handed neutrinos in the Standard Model of
particle physics could be filled in a fully consistent way by
keV-scale sterile neutrinos describing the DM.

\section{Rhenium 187 beta decay and sterile neutrino mass\label{sec:renio}}
As a probe to detect possible mixing of keV sterile neutrinos with
light active neutrinos, we consider in this section the beta decay of
Rhenium 187 ($^{187}$Re; $Z = 75, A = 187$) into Osmium 187
($^{187}$Os; $ Z = 76, \; A = 187 $),
\begin{equation}
^{187}\text{Re} \:\rightarrow \:^{187}\text{Os} + e^- + \bar{\nu}_{e}
 \label{eq:mix }   
\end{equation}

The neutrino flavor eigenstate $ {\nu}_{e} $ (and equivalently for $
\bar{\nu}_{e} $) can be written as a combination of light active
(subscript $i$) and heavy sterile mass eigenstates as \cite{dk,shr}
\begin{equation}
|\nu_{e}\rangle = \sum_{i} U_{ei} |\nu_{i}\rangle +  \sum_{s} U_{es} |\nu_{s}\rangle \
 \label{eq:mixing1}   
\end{equation}
where the quantities $U$ belong to the unitary leptonic mixing
matrix. For the purpose of this paper, we approximate this combination
as a mixing of two mass eigenstates given by \cite{dk}
\begin{equation}
|\nu_{e}\rangle = \cos\zeta \:|\nu_{l}\rangle + \sin\zeta \:|\nu_{s}\rangle
 \label{eq:mixing2}   
\end{equation}
where $ \zeta $ is the mixing angle between a light neutrino mass
state $\nu_{l} $, and the heavy sterile neutrino mass state $ \nu_{s}
$. Other neutrino mass states will not be taken into account in this
work. An effective mass $ m_{l} $ can be used for the former
combination of light mass active neutrinos, but its value ($ m_{l}
\lesssim $ eV) is negligible in comparison with the sterile neutrino
mass in the keV scale. As for the mixing angle $ \zeta $, the
cosmological constraints based on the observed average DM density
suggest \cite{dk,bb}
\begin{equation} \label{eq:zetarang}
\sin^{2}\zeta \sim 10^{-8} \quad ,  \quad \zeta \sim 0.006^{\text{o}} \quad .
\end{equation}
We should keep in mind that these constraints on the value of $ \zeta
$ depend both on the sterile neutrino model and on the sterile
neutrino production mechanism. Eq. (\ref{eq:zetarang}) corresponds to
currently popular models of DM sterile neutrino \cite{dk,bb}.

$^{187}$Re is a long half-life isotope ($ t_{1/2} \simeq 4.35 \cdot
10^{10} $ years), with ground state spin-parity assignment $ J^{\pi} =
5/2^{+} $, that has a single $\beta^{-}$-decay branch mode to the
ground state $1/2^{-}$ of $^{187}$Os with an endpoint energy $
Q_{\beta} \simeq 2.469 $ keV ($ Q_{\beta} = T_e + m_{\nu} + T_{\nu} $,
where $T_{e}$ and $T_{\nu}$ are kinetic energies of the electron and
the neutrino respectively).

In this transition, the change of total angular momentum is $ \Delta J
= 2 $ and there is also a change of parity ($ \Delta \pi=-
$). Therefore we are dealing at best with a first forbidden
Gamow-Teller process.  The lepton system $ (e-\bar{\nu} ) $ carries an
orbital angular momentum $L = 1$ (first forbidden transition) and a
spin $ S = 1 $ (unique Gamow-Teller transition), that couple to the
total angular momentum $J = 2$. The two possible angular momentum
components of the system, $[(l_j)_{e} (l_j)_{\bar{\nu}} ]_{J=2}$, are
$[(p_{3/2})_{e} (s_{1/2})_{\bar{\nu}} ]_{J=2}$ and $[(s_{1/2})_{e}
(p_{3/2})_{\bar{\nu}} ]_{J=2}$. Therefore, as noted in \cite{dvor},
the total differential decay rate $d\Gamma /dE_e$ is a sum of the two
contributions corresponding to the emission of electrons in $p$-wave
and in $s$-wave
\begin{equation}
\frac{d\Gamma}{dE_{e}} =  \frac{d\Gamma_{p_{3/2}}}{dE_{e}} +  \frac{d\Gamma_{s_{1/2}}}{dE_{e}}
 \label{eq:Gama}   
\end{equation}
Following Eq.~(\ref{eq:mixing2}), we write the theoretical spectral
shape of the electron in an $(l_j)$-wave as a sum of the contributions
from light (l) and sterile (s) neutrinos,
\begin{equation}
\frac{d\Gamma_{l_j}}{dE_{e}} =  \: \frac{d\Gamma^{l}_{l_j}}{dE_{e}} \; 
\cos^{2}\zeta+ \:\frac{d\Gamma^{s}_{l_j}}{dE_{e}} \; \sin^{2}\zeta
 \label{eq:gama_a}   
\end{equation}
where
\begin{equation}
\frac{d\Gamma^{\chi}_{l_j}}{dE_{e}} = C \; B_{Re} \; R^2_{Re} \: p_{e} \: p_{\nu_{\chi}} 
\:E_{e} \: (E_{0} - E_{e}) \: F_0(Z, E_e) \: S_l(p_e,p_{\nu_{\chi}}) \: \theta(E_{0} - E_{e} - m_{\chi}) \; ,
 \label{eq:gama_b}
\end{equation}
for $ \chi = l, s$. $ Z $ stands for the atomic number of the daughter
nucleus, $ F_0(Z, E_e) $ is the Fermi function and $ \theta (E_0 - E_e
- m_{\chi}) $ is the step function. $ R_{Re} $ is the nuclear radius
\cite{bebu}, $ B_{Re} $ is the dimensionless squared nuclear reduced
matrix element (r.m.e.) and $ C $ is a constant to be defined later
on. In the above expression, $ E_{e}, \; E_0 $ and $
p_{e}=\sqrt{E_e^2-m_e^2} $ are the total energy, maximum total energy
and momentum of the emitted electron respectively, and $
p_{\nu}=\sqrt{(E_0-E_e)^2-m_{\nu}^2} $ is the momentum of the emitted
neutrino.

Being $Q_{\beta}$ the endpoint energy, we have $ E_0=m_{e}+Q_{\beta}
$, and the kinematical ranges of $ E_{e}, \; p_{e} $ and $ p_{\nu} $
for zero neutrino mass are as follows
\begin{equation}
m_{e}\leq E_{e} \leq m_{e}+Q_{\beta} ; \;\;\;
0 \leq p_{e} \leq \sqrt{Q^2_{\beta}+2 \, m_{e}Q_{\beta}} ; \;\:\;
0 \leq p_{\nu} \leq Q_{\beta} .
\end{equation}
The shape factor $ S_l(p_e,p_{\nu}) $ appears in forbidden decays. For
the case of interest here, a first forbidden decay, $ l $ takes the
value $ l=0 $ for the $ s$-wave and $ l=1 $ for the $ p$-wave
electrons, with shape factors
\begin{equation}
S_0(p_e,p_{\nu})= \frac13 \: p_{\nu}^{2}   \quad \text{and} \quad S_1(p_e,p_{\nu})=  
\frac13  \: p_{e}^{2} \: \frac{F_{1}(Z, E_{e})}{F_0(Z, E_e)} \; .
 \label{eq:shapefactors}
\end{equation}
The relativistic Fermi functions $ F_{0}(Z, E_{e}) $ and $ F_{1}(Z,
E_{e}) $ account for the Coulomb interaction between the residual
nucleus ($ Z= 76 $ in our case) and the emitted electron in the $ s $
and $ p$-waves respectively. They are defined as
\begin{equation}
F_{k-1} = \left[\frac{\Gamma (2k+1)}{\Gamma (k) \:\Gamma(1+2 \gamma_k)}\right]^2   
(2 \, p_{e} \; R)^{2 \, (\gamma_k - k)}  \mid \Gamma (\gamma_k + iz) \mid^2 e^{\pi z} 
 \label{eq:F}   
\end{equation}
and depend on the strength of the Coulomb interaction, given by
the fine structure constant $ \alpha \simeq 1/137.03 $, through
\begin{equation}
\label{eq:gamak}
\gamma_k = \sqrt{k^{2} - (\alpha Z)^2} \quad \text{and} \quad z = \alpha \; Z \; \frac{E_{e}}{p_{e}}    \; ,
\end{equation}
$ k= $ 1, 2 in our case. We note that the Fermi functions in
Eq. (\ref{eq:F}) satisfy $ F_{k-1}(Z\rightarrow 0, E_{e}) \rightarrow
1 $ for $ \alpha \; Z \rightarrow 0 $ and for any $ k \geq 1 $. The
constant factor $ C $ in Eq.~(\ref{eq:gama_b}) is given by
\begin{equation}
C \equiv \frac{G_{F}^{2} \: V_{ud}^{2} \: c_V^2}{2 \, \pi^3} 
\simeq 2 \times 10^{-36} \; \text{(keV)}^{-4} \; ,
 \label{eq:constantC}
\end{equation}
where $ G_{F} $ is the Fermi constant, $ V_{ud} $ the element of the
Cabibbo-Kobayashi-Maskawa matrix ($ \mid V_{ud} \mid \simeq 0.97 $),
and $ c_V \simeq 1 $ is the strength of the vector charged weak
interaction. The dimensionless squared nuclear reduced matrix element
(r.m.e.), $ B_{Re} $, can be computed directly from the experimental
$^{187}$Re mean-life $ \tau = t_{1/2}/\ln 2 $ as [see
Eq.~(\ref{eq:gama_b})]
\begin{equation}
B_{Re}^{-1} =   \tau \: C \: R^2_{Re} \: \int_{m_e}^{E_0} p_e \: p_{\nu} \: E_e \: (E_0-E_e) \: F_0(Z,E_e) 
\: S(p_e,p_{\nu}) \: dE_e \; ,  \label{eq:B2}   
\end{equation}
and it takes the value $ B_{Re} \simeq 3.6 \times 10^{-4} $ for a value of 
the nuclear radius $ R_{Re} $  approximated
as $ R_{Re} \simeq 1.2 \times (187)^{1/3} \:\text{fm} \simeq 6.86 $ fm.
Microscopic calculations of these quantities are in progress.

\begin{figure*}[]
\centering
\includegraphics[width=0.52\textwidth,angle=270]{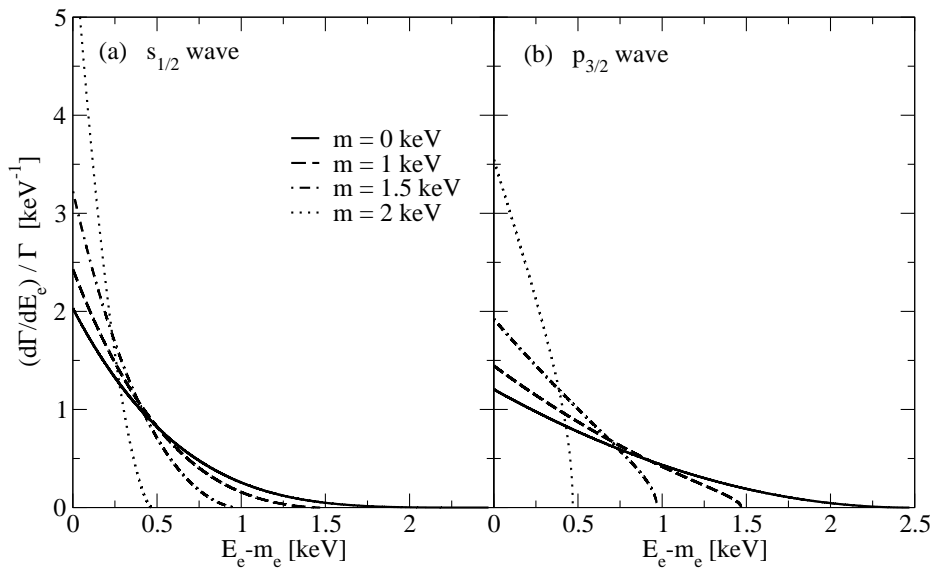}
\caption{Contributions of $s$-wave (left) and $p$-wave (right)
  electrons to the normalized differential decay rate of the process
  $^{187}$Re to $^{187}$Os plotted against the electron kinetic energy
  $ E_e - m_e $. Selected values of the sterile neutrino mass are
  used, $ m_s= $ 1, 1.5, 2 keV (dashed, dashed-dotted and dotted
  lines, respectively), compared to the light neutrino case $ m_l = 0
  $ (solid line).}
\label{fig1}
\end{figure*}

In Fig.~\ref{fig1} we represent the $s$-wave (left) decay rates, $
d\Gamma^{l}_{s_{1/2}}/dE_e $ and $ d\Gamma^{s}_{s_{1/2}}/dE_e $, and
the $p$-wave (right) decay rates $ d\Gamma^{l}_{p_{3/2}}/dE_e $ and $
d\Gamma^{s}_{p_{3/2}}/dE_e $, normalized to one.  We plot the sterile
neutrino contribution for $ s $ and $ p $-wave outgoing electrons and
for selected values of the sterile neutrino mass, $ m= $1, 1.5 and 2
keV (dashed, dash-dotted and dotted line respectively), compared to
the light neutrino case with $m=0$ (solid line).

\begin{figure*}[]
\centering
\includegraphics[width=0.52\textwidth,angle=270]{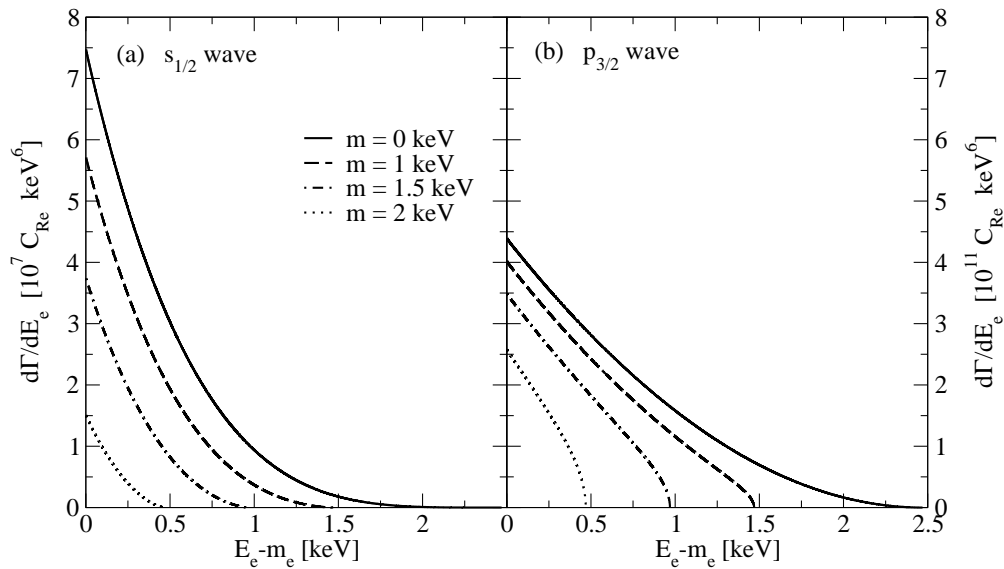}
\caption{Contributions of $s$-wave (left) and $p$-wave (right)
  electrons to the process $^{187}$Re to $^{187}$Os as in
  Fig.~\ref{fig1} but for unnormalized decay rates, in units of
  $10^{7} \:C_{Re}$ keV$^{6}$ (left) and of $10^{11} \:C_{Re}$
  keV$^{6}$ (right). The maximum differential decay rate for $s$-wave
  electrons is of the order of 10$^{7}$, whereas for $p$-wave
  electrons it is 10$^{11}$. The $p$-wave dominates by four orders of
  magnitude for both light and sterile neutrino emission and so the
  spectral shape of beta decay is dictated by the r.h.s. panel.}
\label{fig2}
\end{figure*}

In Fig.~\ref{fig2}, we represent the unnormalized decay rates, on the
left $d\Gamma^{l}_{s_{1/2}}/dE_e$ and $d\Gamma^{s}_{s_{1/2}}/dE_e$,
and on the right $d\Gamma^{l}_{p_{3/2}}/dE_e$ and
$d\Gamma^{s}_{p_{3/2}}/dE_e$. The same choices of neutrino masses as
in Fig.~\ref{fig1} are considered.  As seen in the plots, the maximum
differential decay rate for $s$-wave electrons is of the order of
10$^{7}$ whereas for $p$-wave electrons it is 10$^{11}$ (in units of
$C_{Re}$ keV$^6$ $\equiv C \; B_{Re} \; R_{Re}^2$ keV$^6$). This
dominance by four orders of magnitude of the $p$-wave that was noticed
both theoretically \cite{dvor} and experimentally \cite{AR} for the
light neutrino emission, holds also for sterile neutrino
emission. This is why the spectral shape of beta decay is dictated by
the curves shown in the r.h.s. panel of Fig.~\ref{fig2}.

\begin{figure*}[]
\centering
\includegraphics[width=0.52\textwidth,angle=270]{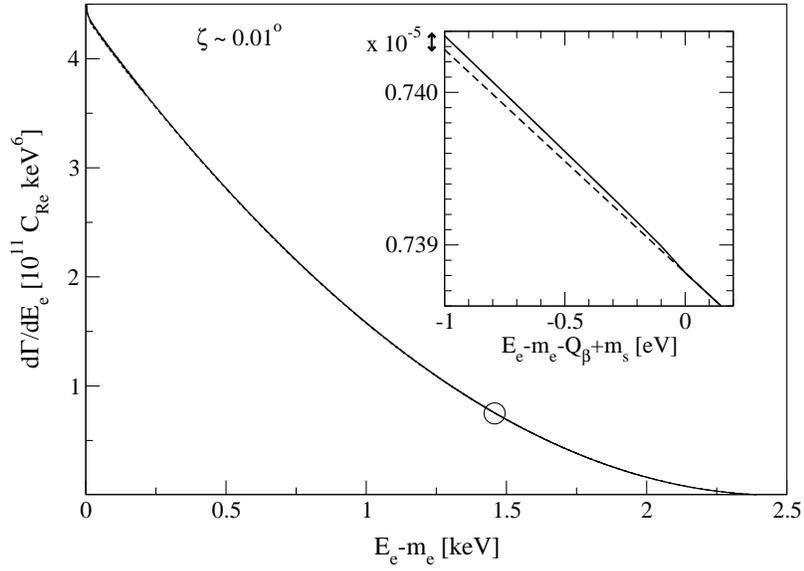}
\caption{The $^{187}$Re to $^{187}$Os beta particle spectrum in units
  of $10^{11} \:C_{Re}$ keV$^{6}$, for a mixing angle $\zeta =$
  0.01$^{\text{o}}$ and sterile mass $m_s= 1$ keV. The curves with
  (solid) and without (dashed) sterile neutrino contribution are
  indistinguishable in the main plot, but are shown in the inset as a
  function of $E_e-m_e-Q_{\beta}+m_s$ in eV, with a separation between
  them magnified as indicated.}
\label{fig3}
\end{figure*}

The effect of the sterile neutrino emission on the electron spectral
shape is represented in Fig.~\ref{fig3} by comparing the differential
decay rate $ d\Gamma/dE_e $ for $ m = 1 $ keV with (solid line) and
without (dashed line) neutrino contribution for $\zeta=$
0.01$^{\text{o}}$. The two curves start to deviate at the step point $
T_e = Q_{\beta} - m_s = 1.469 $ keV, where the sterile neutrino starts
to contribute, and the difference grows as $ T_e $ goes to zero. The
region where the kink appears is shown in the inset on a magnified
scale as a function of $E_e-m_e-Q_{\beta}+m_s$ in eV. The chosen value
of the mixing angle (0.01$^{\text{o}}$) is inspired on Fig. 8 in
Ref. \cite{raX2}, where a plot is made of upper bounds for the mixing
angle as a function of the sterile neutrino mass based on X-ray
observations of dwarf spheroidal galaxies. For different values of the
mixing angle the effect scales as the function ${\mathcal R}$ that we
define below.

In order to analyze the possible effect of a sterile neutrino, we
introduce the dimensionless function
\begin{equation}
 {\mathcal R} \equiv
  \:\frac{d\Gamma^{s}/dE_{e}}{d\Gamma^{l}/dE_{e}}  \tan^2\zeta 
\label{eq:ratioR}
\end{equation}
which is the ratio between the sterile and light neutrino contributions
to the total decay rate times the tangent square of the mixing angle. 
The function $ {\mathcal R} $ is largest for $ p_e $ (or $ T_e $) going to
zero. This procedure is
useful because we are comparing two regions of the same spectrum: the
region where $ (E_e-m_e) < (Q_{\beta}-m_{s}) $ and the emitted
neutrino has enough energy for the sterile neutrino ($ m_s \sim $ keV)
to imprint an effect on the spectrum, and the region where $ (E_e-m_e)
> (Q_{\beta}-m_s) $ and the sterile neutrino effect does not show
up. Clearly, there is a \textit{step} in the spectrum for $ E_e-m_e =
Q_{\beta}-m_s$ which could be observed if the experimental relative
error in this energy region is lower than the height of the step.

\begin{figure*}[]
\centering
\includegraphics[width=0.52\textwidth,angle=270]{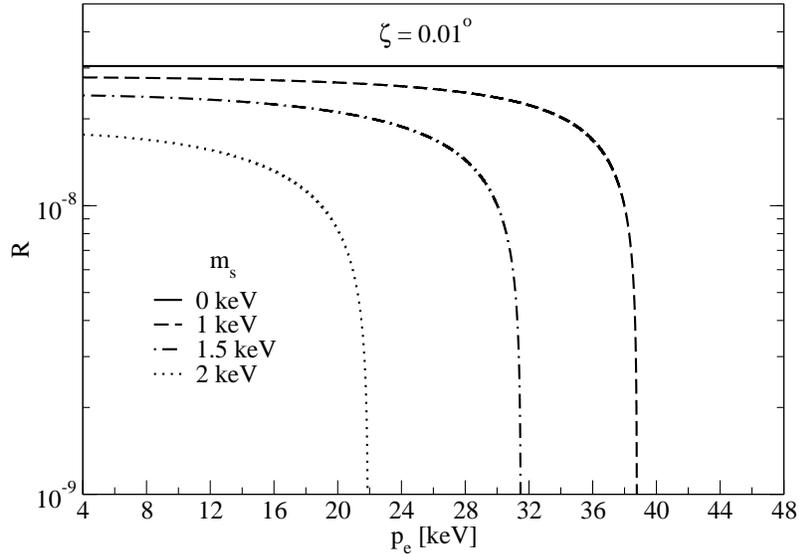}
\caption{Ratio $ {\mathcal R} $, Eq.~(\ref{eq:ratioR}), of the sterile
  neutrino to the active neutrino contributions of the process
  $^{187}$Re to $^{187}$Os vs. the electron momentum for a fixed
  mixing angle $ \zeta= 0.01^{\text{o}} $ and different sterile
  neutrino masses, $ m_s=$ 0, 1, 1.5 and 2 keV (solid, dashed,
  dashed-dotted and dotted lines, respectively). $ {\mathcal R} $
  increases with decreasing $m_s$. ${\mathcal R} $ is nonzero in a
  range $ 0 < p_e < p_{max} $ and $ p_{max} $ decreases as $m_s$
  increases.}
\label{fig4}
\end{figure*}

The ratio $ {\mathcal R} $, Eq.~(\ref{eq:ratioR}), is shown in
Fig.~\ref{fig4} as a function of the electron momentum $p_e$ for a
mixing angle $\zeta = 0.01^{\text{o}}$, and for different values of
the neutrino masses, $m_s=$ 0, 1, 1.5 and 2 keV, corresponding to the
solid, dashed, dash-dotted and dotted lines respectively.  As can be
seen in this figure, the ratio is different from zero in the range $ 0
\leq p_e < (p_e)_{\text{max}} $, with $ (p_e)_{\text{max}} =
     [(Q_{\beta}-m_{s})(Q_{\beta}-m_{s}+2m_e)]^{1/2} $.  For example,
     for $m_s = 2$ keV, $(p_e)_{max} \simeq$ 21.9 keV, and for $m_s =
     1$ keV, $(p_e)_{max} \simeq$ 38.8 keV.  Notice that when $m_s$
     increases, ${\mathcal R}$ decreases.

\begin{figure*}[]
\centering
\includegraphics[width=0.52\textwidth,angle=270]{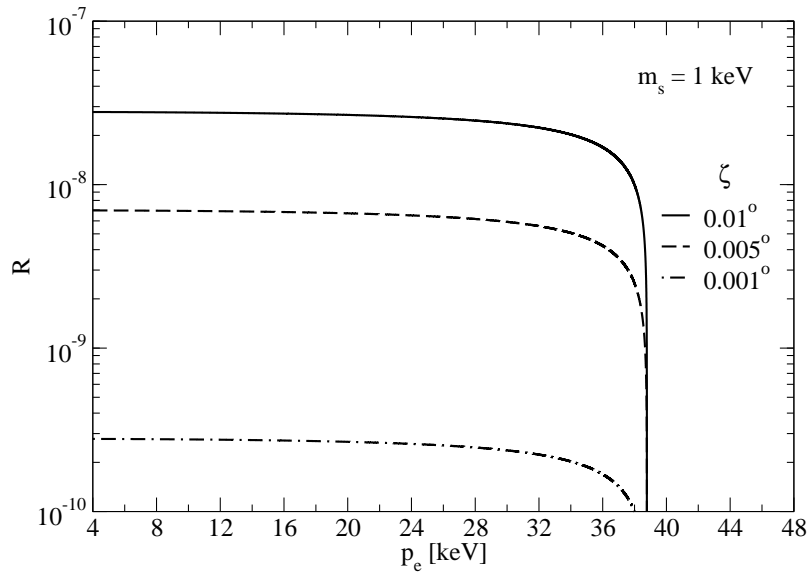}
\caption{The ratio ${\mathcal R}$ as in Fig.~\ref{fig4} but for a
  fixed sterile neutrino mass $m_s=$ 1 keV and different mixing
  angles, $\zeta = 0.01^\text{o} , \; 0.005^\text{o}, \;
  0.001^\text{o}$ (solid, dashed, dashed-dotted lines, respectively)
  of the process $^{187}$Re to $^{187}$Os. ${\mathcal R}$ is almost
  constant in the range $ 0 < p_e < p_{max} $ and increases with
  $\zeta$.}
\label{fig5}
\end{figure*}

Similarly, in Fig.~\ref{fig5} we show the ratio $\mathcal R$ as a
function of the electron momentum $p_e$ but for a fixed sterile
neutrino mass of $m_s =$ 1 keV and different light-sterile mixing
angles $\zeta = 0.01^{\text{o}}, \; 0.005^{\text{o}}, \;
0.001^{\text{o}}$. Figure \ref{fig5} shows not only the increase of
$\mathcal R$ with increasing mixing angle for a fixed value of $m_s$,
but also shows the fact that for fixed values of $m_s$ and $\zeta$,
the ratio $\mathcal R$ is almost constant in the region $ 0 < p_e <
(p_e)_{\text{max}} $.

From Eq.~(\ref{eq:Gama}) and Eq.~(\ref{eq:gama_a}), we write
\begin{equation}
\frac{d\Gamma}{dE_e} = \frac{d\Gamma^{l}}{dE_e}  \:[1 +  {\mathcal R} ]\:  \cos^2\zeta
 \label{eq:GamaR} \; ,
\end{equation}
where $ d\Gamma /dE_e $ is the total differential decay rate when
 neutrino mixing is present ($ \zeta \neq 0 $),
and $ d\Gamma^{l} /dE_e $ is the differential decay rate for the light
neutrino (no mixing: $ \zeta = 0 $).  For small mixing angle $ \zeta $,
the differential decay rate $ d\Gamma /dE_e $  [Eq.~(\ref{eq:GamaR})]
normalized to $ d\Gamma^{l} /dE_e $ is 
\begin{equation}
\frac{d\Gamma/ dE_e}{d\Gamma^{l} /dE_e} \simeq 1 +   {\mathcal R} \; ,
 \label{eq: Gamanorm}
\end{equation}
which shows that for small mixing angle the ratio between the
differential decay rates with mixing and without mixing is given by $
1 + {\mathcal R}$.  This ratio is larger for $ p_e $ or $ T_e $ going
to zero.

We want to emphasize that the energy region suitable for creation and
detection of the keV sterile neutrino corresponds to low $ p_e $ or $
T_e $. On the contrary, information on active neutrinos should be
obtained from the region of $ T_e $ close to the endpoint energy $
Q_{\beta} $.
 
From eqs.~(\ref{eq:Gama}), ~(\ref{eq:gama_b}) and
~(\ref{eq:shapefactors}), we can write for the function $ {\mathcal R} $ the
explicit expression

\begin{equation}
{\mathcal R} = \frac{p^3_{\nu_s}}{p^3_{\nu_l}} \; \theta(Q_{\beta} - T_e - m_s)
 \: \frac{1 + \displaystyle \frac{p^2_e}{p^2_{\nu_s}} \; \displaystyle\frac{F_1(Z,E_e)}{F_0(Z,E_e)}}{
 1 + \displaystyle\frac{p^2_e}{p^2_{\nu_l}} \; \displaystyle\frac{F_1(Z,E_e)}{F_0(Z,E_e)}} \; \tan^2{\zeta} 
 \label{eq:Rshapef}   
\end{equation} 

In order to analyze the ratio $ p^2_e \; F_1(Z,E_e) / F_0(Z,E_e) $, it is
worth to define $ F_{k-1}(Z, E_e) $ as

\begin{equation}
F_{k-1}(Z, E_e) \equiv C_{k-1} \: d_{k-1} \: \left (\frac{m_e}{p_e}\right)^{2 \, k-1} \left (\frac{E_e}{m_e}
\; \alpha \; Z \right)^{2 \, \gamma_k - 1}
 \label{eq:Fnew}
\end{equation}

where
\begin{equation}
C_{k-1} \equiv 2 \, \pi \: (2 m_e R)^{2 \, (\gamma_k - k)} 
\left[\frac{\Gamma(2k+1)}{\Gamma (k) \; \Gamma(1+2 \, \gamma_k)} \right]^2
 \; ; \:\:\:
d_{k-1}\equiv \frac{1}{2 \pi} \left (\frac{E_e}{p_e} \; \alpha \; Z \right)^{1- 2 \, \gamma_k}  
\: \left| \Gamma \left(\gamma_k + i \; \alpha \; Z \; \frac{E_e}{p_e} \right) \right|^2
\; \displaystyle e^{\pi \; \alpha \; Z \; \frac{E_e}{p_e}} 
\label{eq:Ck}
\end{equation}

$ \gamma_{k} $ is defined by Eq.~(\ref{eq:gamak}) and $ d_{k-1}(\alpha
\; Z \; E_e/ p_e) \rightarrow 1 $ for $ p_e \rightarrow 0 $ (for all $
k $). The above definitions yield again $ F_{k-1} \rightarrow 1 $ for
$ \alpha \; Z \rightarrow 0 $ (for all $ k $). In this respect,
eqs.~(\ref{eq:Fnew}) and ~(\ref{eq:Ck}) differ from references
\cite{dvor} and \cite{doko,dokota}.  Finally, the ratio in the shape factor
of Eq.(\ref{eq:shapefactors}) for $ l = 1 $ is given by

\begin{equation}
\frac{p^2_e \: F_1(Z, E_e)}{F_0(Z, E_e)} = \frac{C_1}{C_0} \; \frac{d_1}{d_0} \; m^2_e 
\; \left(\frac{E_e}{m_e} \; \alpha \; Z \right)^{2 \, (\gamma_2 - \gamma_1)}
 \label{eq: quotientpF}
\end{equation}

and for $ p_e \rightarrow 0 $
\begin{equation}
\frac{p^2_e F_1(Z, E_e)}{F_0(Z, E_e)} \buildrel{ p_e \rightarrow 0}\over\sim 0.14 ~ m^2_e \; ,
 \label{eq:quotient0}
\end{equation}
$ [ d_1 / d_0 \rightarrow 1 $ for $ p_e \rightarrow 0 ]$. 

From Eq.~(\ref{eq:Rshapef}) and Eq.~(\ref{eq:quotient0}), we have 

\begin{equation} \label{eq: quotientps}
{\mathcal R} \buildrel{ p_e \rightarrow 0}\over\sim \frac{p_{\nu_s}}{p_{\nu_l}} \; \tan^2{\zeta} \; .
\end{equation}

For $ p_e \rightarrow 0 $ the maximum neutrino momenta are: $
(p_{\nu_s})_{\max} \sim 1.45 $ keV, $ 2.26 $ keV for $ m_s \simeq 2 $
keV, $ 1 $ keV respectively; and $ (p_{\nu_l})_{max} = 2.469 $ keV for
$ m_l = 0 $ keV.

We have shown the relevance of the function $ {\mathcal R} $ to the
analysis of the sterile neutrino effect. One can also study the effect
of a sterile neutrino through the difference between the decay rate
with mixing ($ \zeta \neq 0 $) and the reference case without mixing
($ \zeta = 0 $). This difference is very small, since the mixing is in
any case small, as it is expressed in the following ratio,
\begin{equation}
 {\mathcal R}^* = \displaystyle \frac{\displaystyle\left[\frac{d\Gamma}{dE_{e}}\right]_{\zeta \neq 0}-
\left[\frac{d\Gamma}{dE_{e}}\right]_{\zeta = 0}}{\displaystyle\left[\frac{d\Gamma}{dE_{e}}\right]_{\zeta = 0}} 
= \left(-1+\frac{d\Gamma^{s}/dE_{e}}{d\Gamma^{l}/dE_{e}}\right) \sin^2\zeta
 = \frac{d\Gamma/dE_e}{d\Gamma_l/dE_e} - 1  \; ,
\label{ratioR*}
\end{equation}
which can be written as well as a function of ${\mathcal R}$
\begin{equation}
 {\mathcal R}^*  = - \sin^{2}\zeta + {\mathcal R} \cos^{2}\zeta
 \label{ratioR*R}
\end{equation}

\begin{figure*}[]
\centering
\includegraphics[width=0.52\textwidth,angle=270]{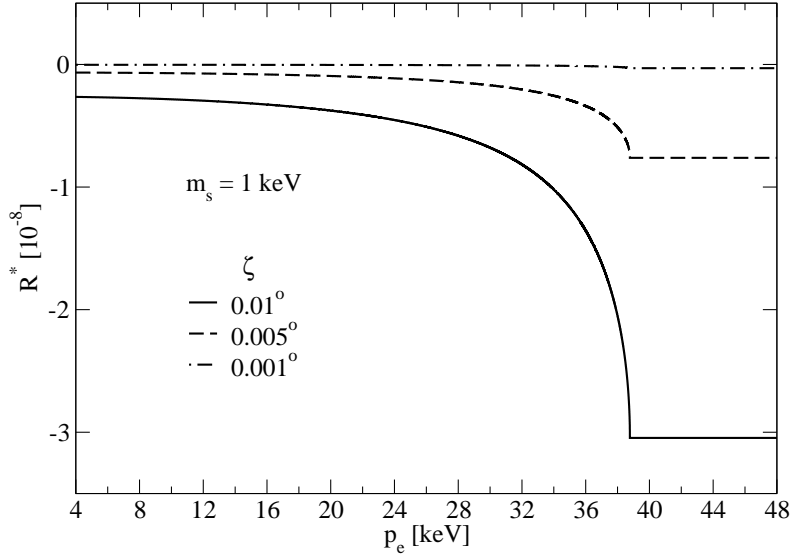}
\caption{Same as in Fig.~\ref{fig5} but for ${\mathcal R^*}$,
  Eq.~(\ref{ratioR*} or Eq.~(\ref{ratioR*R}), in units of $10^{-8}$
  for fixed $m_s$ and different mixing angles of the process $^{187}$Re
  to $^{187}$Os.  The small $p_e$ region is always the best for the
  sterile neutrino detection.}
\label{fig6}
\end{figure*}

In Fig.~\ref{fig6} we plot the quantity ${\mathcal R}^*$ for a fixed
sterile mass $m_s = 1$ keV and for different mixing angles.

The Kurie function is defined as
\begin{equation}
K(y) = \sqrt{\frac{d \Gamma/ dE_e}{p_e \; E_e \; F_0 (Z, E_e) \; S_1(Z, E_e)}}
\quad ,\quad  y \equiv E_0 - E_e = Q - T_e \geq 0 \quad .
\label{eq: K}   
\end{equation}
Considering the mixing
between the light and sterile neutrinos, $K(y)$ can be written as
\begin{equation}
K(y) = \sqrt{ K^2_l(y) \; \cos^2\zeta +  K^2_s (y) \; \sin^2\zeta}
 \label{eq: Kmixing}   
\end{equation}
where
\begin{equation} \label{eq:Kchi}   
K_{\chi}(y) = 
\sqrt{\frac{d \Gamma^{\chi}/ dE_e}{p_e \; E_e \; F_0 (Z, E_e) \; S_1(Z, E_e)}  }
 \quad , \quad \chi = l, \; s \quad .
\end{equation}
For $ \zeta = 0 $ (no sterile-light neutrino mixing), and due to the
introduction of $ F_0 (Z, E_e)\: S_1(Z, E_e) $ in the denominator, $
K(y) $ vs. $ y $ is a straight line for $ m_l \simeq 0 $. This follows
straightforwardly from Eq. (\ref{eq: K}) for $ K(y) $ and from
Eq. (\ref{eq:gama_b}) for $ d \Gamma/ dE_e $, setting $ m_l = 0 $ and $
p_\nu = E_0 - E_e = y $, and it therefore follows that $ K(y) \simeq
{\rm const} \times y $.
$ K(y) $ can be written as well 
in terms of $ \mathcal R $ Eq.~(\ref{eq:ratioR}),
\begin{equation}
K(y) = K_l(y)  \: \sqrt{1+{\mathcal R}} \: \cos\zeta \; .
 \label{eq:KR}   
\end{equation}

\begin{figure*}[]
\centering
\includegraphics[width=0.52\textwidth,angle=270]{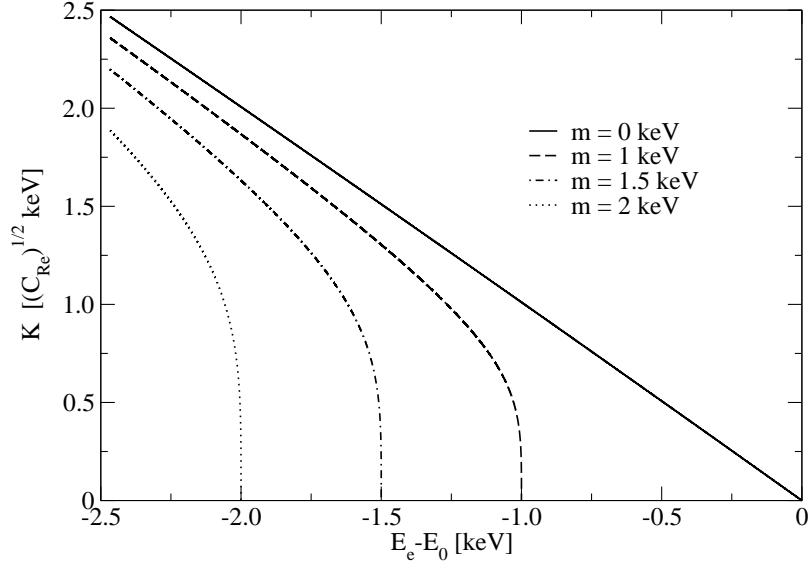}
\caption{Kurie plot $K$ of the process $^{187}$Re to $^{187}$Os for
  different neutrino masses, $ m= $ 0, 1, 1.5 and 2 keV.}
\label{fig7}
\end{figure*}

Finally, in Fig. \ref{fig7} we present the Kurie plot $K$ considering
several neutrino masses, and $ K_l $ (solid line).

\section{Tritium beta decay and sterile neutrino mass}\label{sec:tritium}

Let us now consider the beta decay of Tritium ($^{3}$H; Z = 1; A = 3)
\begin{equation}
^{3}\text{H} \:\rightarrow \:^{3}\text{He} + e^- + \bar{\nu}_{e}
 \label{eq: tritio}   
\end{equation} 
as a probe to detect a possible mixing of keV sterile neutrinos with
active neutrinos. Tritium beta decay would allow the detection of
sterile neutrinos heavier than in the Rhenium beta decay case, within
the 1 to 10 keV range suggested by cosmological and galactic
observations. Tritium, $^{3}$H, is a hydrogen isotope going to the
helium isotope $^{3}$He, with a half-life $t_{1/2} \simeq 12.33 \text{
  years}$, endpoint energy $ Q_{\beta}\simeq 18.59$ keV, and a
spin-parity transition $ 1/2^{+} \rightarrow 1/2^{+} $.

For the Tritium decay case there is no change in angular momentum and
parity corresponding to an allowed transition ($ L = 0 $) with Fermi
($ S = 0 $) and Gamow-Teller ($ S = 1 $) components. Therefore, the
electron is emitted in $s$-wave and the differential decay rate
is simply
\begin{equation}
\frac{d\Gamma}{dE_e} = \frac{d\Gamma_{s_{1/2}}}{dE_e}
 \label{eq:detritio1}   
\end{equation}  
Similarly to Eq.~(\ref{eq:gama_a}) and Eq.~(\ref{eq:gama_b}) we have  
 \begin{equation}
\frac{d\Gamma_{s_{1/2}}}{dE_e} =  \:\frac{d\Gamma^{l}_{s_{1/2}}}{dE_{e}} \; \cos^{2}\zeta
+ \:\frac{d\Gamma^{s}_{s_{1/2}}}{dE_{e}} \; \sin^{2}\zeta
 \label{eq:detritio2}   
\end{equation} 
and
\begin{equation}
\frac{d\Gamma^{\chi}_{s_{1/2}}}{dE_e} =  C B_{T}  \:p_{e} \:p_{\nu_{\chi}} \:E_{e} \:(E_{0} - E_{e}) \:
F_0(Z, E_e)  \:\theta(E_{0} - E_{e} - m_{\chi}) \quad , \quad \chi = l, s \quad ,
 \label{eq:detritio3}   
\end{equation}  
as the shape factor $ S(p_e, p_{\nu})$ is 1 for allowed decays. The
relativistic Fermi function $F_0(Z, E_e)$ was defined in
Eq.~\ref{eq:F}. The squared r.m.e. for the allowed decay of Tritium (T) is
$B_{T} = B_{F_{T}} + B_{GT_{T}}$, where $ B_{F} $ and $ B_{GT} $ are
the Fermi and Gamow-Teller decay strengths respectively, given by: 
\begin{equation}
B_{F_T}  =   \frac{1}{2} \:|\langle\: ^{3}\text{He}(1/2^{+}) \parallel \sum_{j=1}^{A=3}   
\tau_{j}^{+}  \parallel\: ^{3}\text{H}(1/2^{+}) \:\rangle|^2
\label{eq:BF}   
\end{equation}
and
\begin{equation}
B_{GT_T}  =  \frac{g^2_A}{2} \:|\langle\: ^{3}\text{He}(1/2^{+}) \parallel \sum_{j=1}^{A=3}   
\tau_{j}^{+}  \vec{\sigma}_{j}  \parallel\: ^{3}\text{H}(1/2^{+}) \:\rangle|^2 
\label{eq:BGT}
\end{equation}
where $ g_A=c_A/c_V \simeq 1.26 $ is the axial-to-vector strength
ratio of the charged weak interaction.

\begin{figure*}[]
\centering
\includegraphics[width=0.52\textwidth]{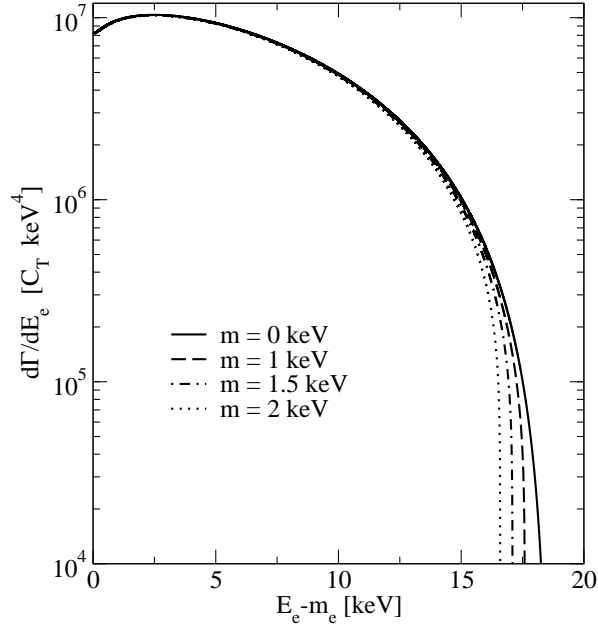}
\caption{Differential decay rate for the process $^{3}$H going to
  $^{3}$He with the emission of active ($ m= $ 0) or sterile ($ m= $
  1, 1.5, 2 keV) neutrinos, in units of $C_T$ keV$^{4}$. The vertical
  axis is in logarithmic scale.}
\label{fig8}
\end{figure*}

\begin{figure*}[]
\centering
\includegraphics[width=0.52\textwidth]{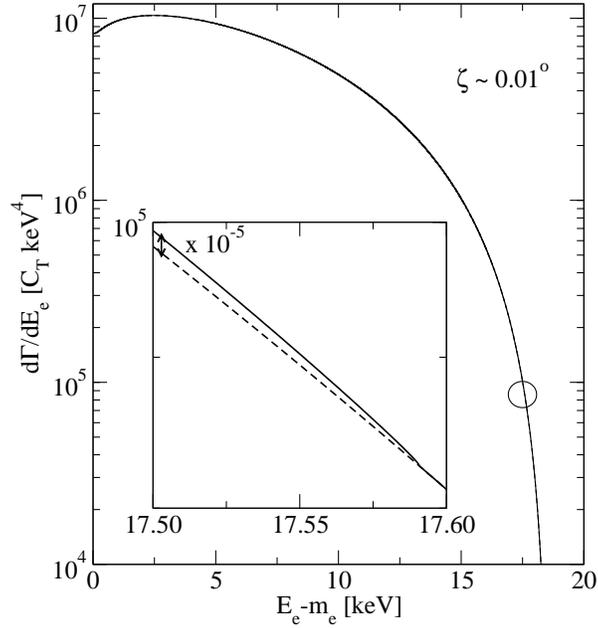}
\caption{The $^{3}$H to $^{3}$He beta particle spectrum with (solid
  line) and without (dashed line) sterile neutrino contribution for
  sterile mass $m_s= 1$ keV and $\zeta =$ 0.01$^{\text{o}}$. Axis labels
  of the inset are the same as in the main plot, where a magnification
  of the separation between curves is magnified as indicated.}
\label{fig9}
\end{figure*}

\begin{figure*}[]
\centering
\includegraphics[width=0.52\textwidth,angle=270]{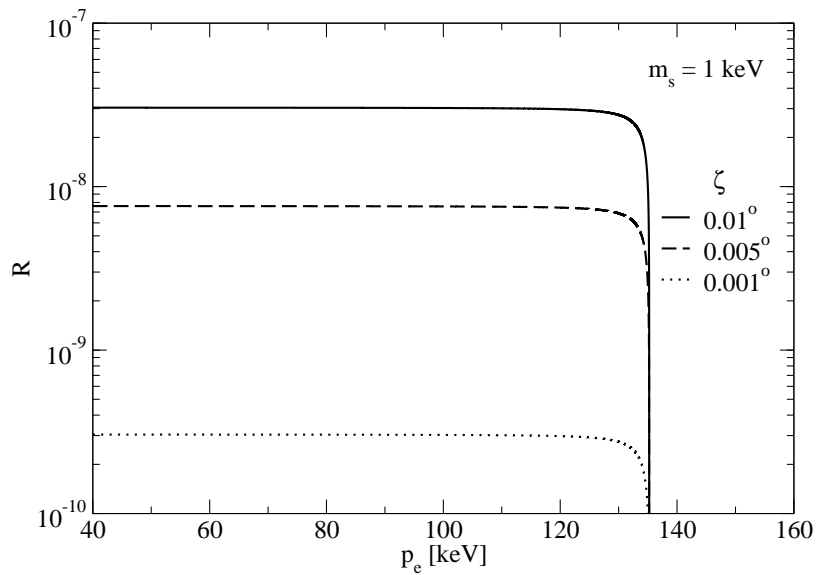}
\caption{Ratio ${\mathcal R}$, Eq.~(\ref{eq:ratioR}), vs. the electron
  momentum for a fixed sterile neutrino mass $m_s=$ 1 keV and
  different mixing angles, for the process $^{3}$H going to $^{3}$He.}
\label{fig10}
\end{figure*}

From the experimental mean-life of Tritium the decay strength can be
obteined from
\begin{equation}
  B_{T}^{-1} =   \tau \: C \: \int_{m_e}^{E_0} p_e \: p_{\nu} \: E_e \: (E_0-E_e) \: F_0(Z,E_e) \: dE_e \; ,  \label{eq:detritio4}   
\end{equation}
which yields a value $ B_T \simeq 5.61 $. In Fig.~\ref{fig8} we plot
the differential decay rates of the process $^{3}$H to $^{3}$He for
neutrinos of different masses. As an illustration of the heavy
neutrino contribution to the total decay rate of the Tritium beta
decay, we plot the electron spectrum in Fig.~\ref{fig9} with (solid
line) and without (dashed line) sterile neutrino contribution with
mass $m_s = 1$ keV and mixing angle 0.01$^{\text{o}}$. The effect is made
visible in the inset thanks to the indicated magnification. Finally,
in Fig.~\ref{fig10} we plot the ratio ${\mathcal R}$ vs. $p_e$ for a
fixed sterile neutrino mass ($m_s = 1$ keV) and different mixing
angles.

\section{Conclusions \label{sec:conclu}}

The detection of sterile neutrinos is not only important from the
point of view of particle physics for the extension of the SM, but
also from the point of view of cosmology and astrophysics as a serious
candidate for dark matter in the keV mass range. With the relevance of
the possible detection of keV scale dark matter candidates in mind, we
have studied Rhenium 187 and Tritium beta decays. The low electron
energy domain of the beta spectrum is the region where a sterile
neutrino could be detected and its mass measured, the expected mass
being in the keV scale (1 to 10 keV) as constrained from cosmological
and galactic observations and theoretical analysis. The electron low
energy region that is suitable to detect the sterile neutrino, $ 0
\lesssim T_e \lesssim (Q_{\beta} - m_s) $, is away from the endpoint
energy region suitable for the detection of the active neutrino mass.

Two experiments are running at present, MARE and KATRIN, dealing with
Rhenium 187 and Tritium beta decays respectively. The MARE experiment
will provide the entire shape of the electron differential decay rate,
giving data for both the sterile and the active neutrino detection
regions. KATRIN so far concentrates on the region near the endpoint of
the electron spectrum but it would be extremely interesting to look
for data in the region where keV sterile neutrinos could show up
\cite{cw}. In this paper, we have carried out the study of the role of
sterile neutrinos in beta decay spectra, within the expected keV mass
range and considering different mixing angles, according to
astronomical and cosmological observations and experiments.

For $^{187}$Re the electrons can be emitted in $p$-wave and in
$s$-wave, the former dominating the decay by a factor $10^4$ over the
latter. For Tritium the electrons are emitted in $s$-wave only. The
spectra of the electrons emitted in these waves have been carefully
computed from the experimental beta decay half-lives using
relativistic Fermi functions. Results for different neutrino masses
(light and in the 1-2 keV range) have been obtained separately and
mixed. We have also computed for both decays the ratio ${\mathcal R}$
of the light to the heavy component of the mixing. It is different
from zero in an electron momentum range $0 \leq p_e <
(p_e)_{\text{max}}$, where $(p_e)_{\text{max}}$ decreases with
increasing sterile neutrino mass $m_s$ (equivalently for electron
kinetic energy range). In the vicinity of $(p_e)_{\text{max}}$ the
ratio ${\mathcal R}$ drops off sharply, but in the rest of the range
it exhibits an almost constant plateau with a slight increase as $p_e$
goes to zero. It increases with the mixing angle and decreases with
the sterile neutrino mass $m_s$.

In order to detect the small deviation in the experimental spectrum
due to the sterile neutrino mixing, the relative experimental random
error (inversely proportional to the square root of the number of
measured events, $ \epsilon \sim N_{\beta}^{-1/2} $) must be as small
as possible. To this end, the number of detected events $N_{\beta}$
must increase by choosing, for instance, a beta decay with a small
$Q_{\beta}$ value or by increasing the time of data acquisition. For
MARE, the typical number of events is $ 10^{13}-10^{14} $ for 10 years
of data acquisition, 8 arrays and 400 gr of natural Rhenium
\cite{mar}. We found that at its largest value, the ratio ${\mathcal
  R}$ of the sterile neutrino to the active neutrino contributions is
about $10^{-8}$ using a realistic mixing angle. Therefore the sterile
neutrino probability ${\mathcal R} \times N_{\beta}$ is about
$10^{5}-10^{6}$, which is not negligible. It implies finding
$10^{5}-10^{6}$ sterile neutrinos within $10^{13}-10^{14}$
events. These numbers increase one order of magnitude for the MARE
option of $10^{15}$ events for 10 years of data acquisition, 16 arrays
and 3.2 kg of natural Rhenium \cite{mar}. A simple estimate requires
the Poisson error $\epsilon$ to be smaller than the ratio ${\mathcal
  R}$. Namely, $N_\beta > 1/{\mathcal R}^2 \sim 10^{14}-10^{15}$. Of
course, in order to assess a precise prediction of the detection
probability one should include a careful analysis of the systematic
errors and instrument parameters, but such study goes beyond the scope
of the present paper. The small effect expected on the electron
spectrum calls for sources with larger stability to reduce the
systematic errors, which pose at present a difficult challenge on the
detection capabilities of these experiments. Furthermore, for
${\mathcal R}=$ 10$^{-8}$ there would be one sterile neutrino event
for one hundred million active neutrino events.

The main purpose of this paper has been to guide future experimental
searches for sterile neutrinos.  From the point of view of particle
physics, one is talking about an extension of the Standard Model.
From the point of view of cosmology, one is looking for a keV
candidate for DM (mass range favoured by cosmological
observations). Therefore, we show in this paper the relevant energy
range where experimentalists should focus on as well as the order of
magnitude of the expected signal, both in absolute terms and with
respect to the background.

\acknowledgments

We are grateful to Peter Biermann, Angelo Nucciotti (MARE) and
Christian Weinheimer (KATRIN) for useful discussions. O.M and E.M.G
acknowledge the Spanish Ministry of Science and Education for partial
financial support (FIS 2008-01301, FIS 2011-23565 and FPA 2010-17142).
M.R.M. acknowledges the financial support of the Spanish Ministry of
Science and Education (FIS 2008-01323), and the kind hospitality of
the Observatoire de Paris.

\end{document}